# The Case for Cloud Service Trustmarks and Assurance-as-a-Service

Theo Lynn, Philip Healy, Richard McClatchey, John Morrison, Claus Pahl and Brian Lee
*Irish Centre for Cloud Computing & Commerce, Dublin, Ireland*
*theo.lynn@dcu.ie, p.healy@cs.ucc.ie, richard.mcclatchey@uwe.ac.uk, j.morrison@cs.ucc.ie,
cpahl@computing.dcu.ie, blee@ait.ie*

Abstract: Cloud computing represents a significant economic opportunity for Europe. However, this growth is threatened by adoption barriers largely related to trust. This position paper examines trust and confidence issues in cloud computing and advances a case for addressing them through the implementation of a novel trustmark scheme for cloud service providers. The proposed trustmark would be both active and dynamic featuring multi-modal information about the performance of the underlying cloud service. The trustmarks would be informed by live performance data from the cloud service provider, or ideally an independent third-party accountability and assurance service that would communicate up-to-date information relating to service performance and dependability. By combining assurance measures with a remediation scheme, cloud service providers could both signal dependability to customers and the wider marketplace and provide customers, auditors and regulators with a mechanism for determining accountability in the event of failure or non-compliance. As a result, the trustmarks would convey to consumers of cloud services and other stakeholders that strong assurance and accountability measures are in place for the service in question and thereby address trust and confidence issues in cloud computing.

## 1 INTRODUCTION

Surveys of consumers and enterprises have consistently highlighted concerns about entrustment of data to cloud service providers. The extent to which these concerns are exaggerated is open to debate. Regardless, their very existence limits cloud uptake. Section 1.1 reviews the evidence for trust concerns around cloud computing and examines their subjects and scope.

Accountability measures have been proposed as a means of enhancing the trustworthiness of cloud offerings. Although this approach includes preventative aspects, it is characterised by the assigning of liability in the event of failure through detective measures and subsequent remediation, if appropriate. Assurance, in contrast, is concerned with conclusions designed to enhance the degree of confidence of the intended users other than the cloud service provider about the outcome of an evaluation or measurement, typically of dependability and predictability, of a cloud computing service. Section 1.2 provides an overview of accountability and assurance concepts as they apply to cloud computing and considers the similarities and differences between them.

A trustmark is a third-party mark, logo, picture, or symbol that is presented in an effort to dispel consumers' concerns about risk. A well-known example is the VeriSign (now Norton) Trust Seal, which indicates to users of a website that its authenticity and security practices have been independently verified. Section 2 presents the case for trustmarks as a confidence-building measure in the trustworthiness of cloud services. This reasoning is extended in Section 2.1 to include the notion of an "active" and "dynamic" trustmark that incorporate multi-modal display of up-to-date information about the dependability of the underlying service rather than the traditional notion of trustmarks as static and passive badge-like entities.

The utility of a trustmark is dependent on the independent verification procedures that it represents. Section 3 outlines how an assurance service could be implemented that would provide transparency to consumers of cloud services, and other stakeholders, that appropriate assurance and accountability procedures are being followed by a cloud service provider. This information could be surfaced to consumers and other stakeholders via a the revised trustmark concept described in Section 2.1.

We conclude in Section 4 with a summary of the proposed concepts (trustmarks and assurance services) and a discussion of how these could be turned into concrete implementations, and a set of objective measures for success.

## 1.1 Barriers to Cloud Adoption

A recent study carried out by IDC on behalf of DG Connect of the European Commission recently reported that more than half of EU businesses and consumers are using some form of cloud computing service today and suggested cloud computing could contribute up to €250 billion to EU GDP in 2020 and create over 3.8 million jobs (IDC, 2012). Despite these relatively high adoption rates, the same report highlights that cloud adoption by business users is currently higher for "basic" solutions and cloud intensity (measured as the number of cloud solutions adopted by each enterprise) is also rather limited. The impact of these two factors is that the potential value of spending activity on public clouds is reduced by up to 56% (IDC, 2012).

The barriers causing this limited cloud adoption behaviour are well documented and cluster around five areas. Both consumers and enterprises have concerns over the location, integrity, portability, security and privacy of their data in the cloud (HBR Analytics, 2010; Yankee Group, 2010; Pew Internet (2010); IDC (2012)). Regulatory responses to these concerns are resulting in increasing complexity with different compliance requirements across industry sectors and legal jurisdictions. These trust and compliance complexity issues are further exacerbated by the nature of businesses operating in the cloud, often characterised by a chain of service provision and trans-border data transfers. Governance issues, data loss and leakage, and shared technology vulnerabilities are amongst the top threats to cloud computing identified by independent international organisations (Cloud Security Alliance, 2010).

Recently, the European Commission suggested a series of actions to overcome some of these barriers through (i) opening up access to content, (ii) making online and cross-border transactions straightforward and (iii) building digital confidence (European Commission, 2012). With regards to cloud computing, the European Commission is more specific about the need for *"a chain of confidence-building steps to create trust in cloud solutions."* These steps include actions to:

- promote wider use of standards, the certification of cloud services to show they meet these standards and the endorsement of such certificates by regulatory authorities;
- develop model terms for cloud computing service level agreements and standardised contract terms and conditions for various use scenarios including business-to-business, business-to-consumer and trans-border business; and,
- promote common public sector leadership through a European Cloud Partnership to work on common procurement requirements for cloud computing. (European Commission, 2012)

Addressing these challenges is not trivial. Digital business ecosystems built on chains of service provisions, such as those underlying cloud computing, introduce different issues that can undermine trustworthiness and dependability. These include provenance of data and technology, transparency about the operation and implementation of technology, and processes and the predictability of the technology to behave within expected norms (*i.e.*, dependability) (Sommerville, 2007). Mechanisms to instil trustworthiness and dependability in cloud computing may include:

- new approaches to constructing dependability arguments;
- methods and tools for testing cloud infrastructures and configurations;
- self-aware systems that make information about their operation and failure available for scrutiny and use; and,
- regulatory and social mechanisms to highlight dependable and trustworthy service providers; and
- de-emphasise or remove undependable and untrustworthy elements of the chain of service provision (Somerville, 2007).

## 1.2 Accountability and Assurance

Accountability has been proposed as a potential solution for addressing the deficit in trust and confidence in cloud computing (Pearson *et al.*, 2012; Ko *et al.*, 2011; Pearson and Charlesworth, 2009; Haeberlen, 2010). For example, the A4Cloud project proposes that a chain of accountability for data in the cloud can be established to provide users, cloud providers and regulators together to clarify liability and provide greater transparency overall (Pearson *et al.*, 2012). It is envisaged that this will be achieved by a set of mechanisms that may mitigate risk (preventative controls), monitor and identify risk and policy violation (detective controls), manage incidents and provide redress (corrective controls). The projects typically focus on scenarios involving *a priori* contracts between parties and as such do not serve to address trust and confidence issues in the wider marketplace but are restricted to those party to the service level agreement.

Accountability may be defined as: *"The property of a system or system resource that ensures that the actions of a system entity may be traced uniquely to that entity, which can then be held responsible for its actions"* (Huff, 1981 as cited in IETF RFC4949). Assurance is different. It is concerned with conclusions by practitioners designed to enhance the degree of confidence of the intended users other than the responsible party about the outcome of an evaluation or measurement of a subject matter against criteria (adapted from International Framework of Assurance Engagements, IAASB, 2008). One might argue that it is not *ab initio* focussed on failure and liability but rather focussed on dependability and, given sufficient time and repetition, predictability. The benefits of assurance include:

- an independent opinion from an external source that enhances credibility of the cloud service;
- reduction of perceived management bias in service claims;
- improved relevance of information associated with the expertise and knowledge of the assurance provider; and,
- modification of perceived risk.

While accountability in cloud computing is focussed on provenance and transparency, it does not adequately address predictability in the way that assurance can. The natural extension of an assurance-based approach to cloud computing is a Design for Dependability (in contrast to a Design for Failure) model that integrates both fault avoidance and fault tolerance. It should be noted that assurance, unlike accountability, need not be absolute given the limitations and uncertainties inherent in the chain of service provision and the impracticality of examining all evidence associated with any given cloud computing service but it may be reasonable given an accumulation of evidence.

We posit that an integrated approach to accountability and assurance is needed for cloud computing.

A proposed integrated framework is presented in Figure 1 that extends and realigns the Pearson and Wainwright (2012) Framework for Accountability.

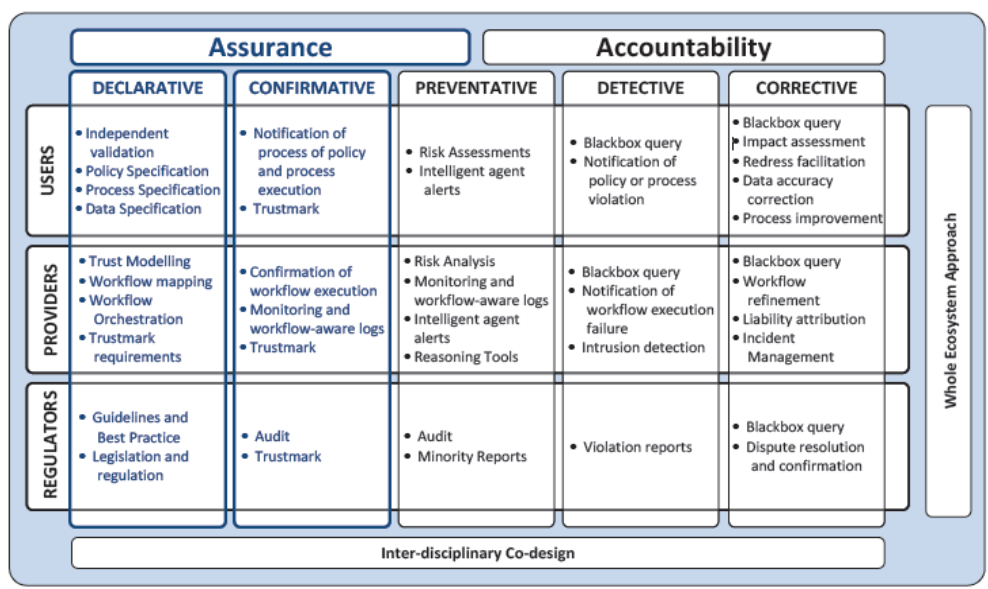

Figure 1 An Integrated Framework for Assurance and Accountability (an adaptation of Pearson and Wainwright, 2012)

## 2. THE CASE FOR CLOUD TRUSTMARKS

The issues around trust raised by the advent of cloud computing are similar to those raised by the Internet as a whole. Consumers of cloud computing, like general Internet consumers, must trust that cloud service providers will not default on implied or explicit bonds, that the service quality is good and will be delivered as promised, and that their personal information will be securely held and their privacy respected (Aiken and Boush, 2006).

Trustmarks are any third-party mark, logo, picture, or symbol that is presented in an effort to dispel consumers' concerns about risk and therefore increase firm-specific trust levels (Aiken and Boush, 2003). Trustmark services typically involve one or more of six elements: (i) a declaration of best practice, (ii) a subscription to a code of conduct, (iii) scrutiny for membership (based on criteria), (iv) sanctions for failure to adhere to a code of conduct, (v) recourse (appeals) for wrongful revocation of the trustmark and (vi) a remedy for aggrieved customers (Endeshaw, 2001). Policymakers, academia and industry have called for research on trustmarks in the cloud computing context (IAMCP, 2011; GAP Task Force, 2011; Robinson *et al.*, 2010). Research suggests that trustmarks have the greatest effect on perceived trustworthiness in an Internet context (when compared to objective source third-party ratings and advertising-derived implications), influencing respondents' beliefs about security and privacy, general beliefs about firm trustworthiness, and willingness to provide personal information (Aiken and Boush, 2006). Recipients of trustmarks are typically subject to a manual verification and certification process that varies widely within the trustmark provision sector and is not transparent to cloud customers and consumers. Trustmarks have been criticised for consistency, reliability, currency / timeliness, accuracy, transparency and ease of abuse (Schouten, 2012; Endeshaw, 2001; Remotti, 2012). Trustmarks are widely used in e-commerce (Remotti, 2012). We posit that existing static passive conceptualisations of trustmarks will not satisfactorily address the trust and confidence issues in cloud computing due to the inherently dynamic nature of these services. As such, we propose an active dynamic trustmark system for cloud computing that overcomes the shortfalls of accountability, assurance and trustmarks as discrete solutions for trustworthiness issues in cloud computing.

### 2.1 Active Dynamic Trustmarks

As noted above, trustmarks are typically presented as a static visual representation, typically a badge-like logo, on websites or promotional material. However, trustmarks need not be static; by utilising modern web technologies, such as HTML 5, trustmarks could be presented as active dynamic entities that succinctly communicate up-to-date values for a number of high-level dependability measures. These dependability measures would be based on "live" analytics of aspects of the underlying service. Static badge-like images could be replaced by multi-modal entities that communicate information (i) graphically using lightweight, standard-compliant technologies such as HTML 5 canvas (ii) textually and (iii) in a machine readable format via semantic web technologies such as OWL. Furthermore, the authenticity of these trustmarks can be verified by a certification mechanism. Unlike the opaque assurance-backed certification approach that has been traditionally applied to trustmarks, active dynamic trustmarks would provide prospective and existing cloud customers with a window into the operation of the underlying cloud service by providing a mechanism that would allow users to "drill down" into specific high-level metrics, at that moment or over a period of time, that comprise the trustmark. As a result, stakeholders can satisfy themselves that the service is both trustworthy and dependable and the level of trustworthiness is signalled to the market as a whole. The design of the trustmark interface would need to balance the need to inform stakeholders with varying roles against privacy and security concerns. Discrete independent virtualised services could be provided for internal and external auditors and regulators to analyse service performance against business policy, legal or regulatory compliance requirements.

## 3. ASSURANCE-AS-A-SERVICE

To deliver the real-time metrics communicated by the active dynamic trustmarks, as proposed in Section 2.1, necessitates the collection, collation and computation of data relating to the operation of the service. These metrics must be re-evaluated on an ongoing basis with the resulting data being communicated to the trustmark metadata platform before being surfaced via multi-modal trustmark updates. This task could be delivered by the cloud service provider or offloaded to an independent third-party assurance service, which could itself be a cloud service. In its primary capacity, the assurance service would watch the operation of other cloud services and surface data to the trustmark interface.

One approach to services monitoring could be the declaration and confirmation of processes that describe the flow of execution of the processes that make up a cloud service. The service being assured would then communicate the commencement and completion of actions via web service calls to a checkpointing mechanism. Checkpointing would serve two purposes: (a) the creation of an independent provenance log of actions performed and their context (who, what, when, where), providing a basis for assurance monitoring and (b) the steering of processes and intervention where appropriate.

If the declared processes were sufficiently fine-grained, then large volumes of heterogeneous data could be created at rates beyond the monitoring and analytical capabilities of cloud consumers. Furthermore, consumers should be notified of dependability events without continuous monitoring of the trustmark. This shortcoming could be addressed by providing a mechanism that would continuously monitor provenance logs on behalf of consumers through the use of software components that represent the interests of the various internal and external, direct and indirect, stakeholders. Integrating these mechanisms with a hive of software agents would ease the monitoring and auditing burden on stakeholders by encoding their rules and ensuring that they are enforced on an ongoing basis. Furthermore, agents could also automatically trigger detective, preventative and remediation activities where appropriate. Monitoring would therefore not simply be about identifying and laying the blame for faults and transgressions (accountability); it would actively try to engage with service providers, consumers and other stakeholders to confirm quality of service and where necessary identify, prevent and resolve issues in a transparent fashion while signalling its effectiveness or otherwise at doing this in a transparent fashion (assurance).

## 4. CONCLUSIONS

Trustmarks have a proven track record in addressing the internet-related trust concerns held by consumers. They therefore may have a role to play in addressing the deficit in trust that has been identified in users' attitude towards cloud services. However, trustmarks need not be static badge-like entities. Modern web technologies allow for the multimodal communication of a range of up-to-date metrics, allowing the trustmark concept to be expanded to become a much richer communications medium. These active dynamic trustmarks could be backed by third party assurance services that communicate dependability metrics to the trustmark platform.

The success of the independent third-party assurance services, as proposed, would depend on their ability to meet the following objectives:

1. enabling cloud service providers to give their consumers appropriate control and transparency over the definition and execution of their processes and workflows within cloud services;
2. assuring cloud consumers that the processes and workflows as defined were executed in accordance with their declared expectations;
3. monitoring and confirming consistent predictable compliance of cloud service providers with their clients' declared expectations, business policies, regulations and third-party standards;
4. providing a mechanism for verifying the integrity of the provenance chain without learning details of individual records; and
5. signalling dependability and trustworthiness of cloud service providers to stakeholders using an agreed set of measures through an objective trustmark service.

In summary, it is posited that a Design for Dependability model of cloud computing, supported by an independent third-party assurance system that recognises and communicates consistent predictable service through an independent active and dynamic trustmark, would improve trustworthiness and confidence in cloud computing and serve to differentiate cloud service providers.